\documentclass[letter]{aa}
\usepackage{amsmath}
\usepackage{graphicx}
\usepackage{txfonts}
\usepackage{natbib}
\bibpunct{(}{)}{;}{a}{}{,} 

\begin{document}

\title{The shear-Hall instability in newborn neutron stars}

\author{T. Kondi\'{c}\inst{1}\and G. R\"udiger\inst{1}\and R. Hollerbach\inst{2}}

\institute{Astrophysikalisches Institut Potsdam, An der Sternwarte 16, D-14482
Potsdam, Germany\\
 e-mail: gruediger@aip.de; tkondic@aip.de \and University of Leeds,
Department of Applied Mathematics, Leeds LS2 9JT, UK\\
 e-mail: rh@maths.leeds.ac.uk }

\date{Received ; accepted }

\abstract{}{In the first few minutes of a newborn neutron star's life the Hall
effect and differential rotation may both be important. We demonstrate that
these two ingredients are sufficient for generating a `shear-Hall instability'
and for studying its excitation conditions, growth rates, and characteristic
magnetic field patterns.}{We numerically solve the induction equation in a
spherical shell, with a kinematically prescribed differential rotation profile
$\Omega(s)$, where $s$ is the cylindrical radius. The Hall term is linearized
about an imposed uniform axial field.  The linear stability of individual
azimuthal modes, both axisymmetric and non-axisymmetric, is then
investigated.}{For the shear-Hall instability to occur, the axial field must be
parallel to the rotation axis if $\Omega(s)$ decreases outward, whereas if
$\Omega(s)$ increases outward it must be anti-parallel. The instability draws
its energy from the differential rotation, and occurs on the short rotational
timescale rather than on the much longer Hall timescale. It operates most efficiently if
the Hall time is comparable to the diffusion time.  Depending on the precise
field strengths $B_0$, either axisymmetric or non-axisymmetric modes may be the
most unstable.}{Even if the differential rotation in newborn neutron stars is
quenched within minutes, the shear-Hall instability may nevertheless amplify
any seed magnetic fields by many orders of magnitude.}


\authorrunning{T. Kondi\'{c}, G. R\"udiger \and R. Hollerbach} \titlerunning{The
shear-Hall instability} 

\maketitle

\section{Introduction}

The classical dynamo mechanism for generating cosmic magnetic fields is based
on the inductive effects of rotating anisotropic turbulence.  Because the same
turbulence prevents uniform rotation from being a solution of the Reynolds
equation, in the majority of dynamo models, the basic mechanism for producing
toroidal fields from poloidal fields is differential rotation. However, it
alone only produces toroidal fields that grow linearly in time, until the
resulting Maxwell stresses suppress the differential rotation. To produce a
self-excitation of magnetic fields as an instability that grows exponentially
in time, another effect is needed to produce poloidal fields from toroidal ones.

There are many possibilities. The simplest one is a meridional flow.  For
certain combinations of differential rotation, meridional flows, and finite
conductivity, the coupled system allows the self-excitation of (purely
non-axisymmetric) fields \citep[e.g. ][]{duja1989}. It is also possible to
close the circuit by the $\alpha$-effect, which represents the existence of an
electromotive force parallel to the mean magnetic field. The result is a
so-called $\alpha\Omega$-dynamo, which for a high magnetic Prandtl number can be
modified by the action of meridional flow \citep{chscdi1995,kueruesc2001}.
Driven turbulence subject to a shear flow in liquid metals is also able to work
as a dynamo if the turbulence intensity is stratified in the direction
orthogonal to the shear-flow plane \citep{rueki2005}.

The dynamo equations resulting from purely hydrodynamic flows are
all linear, and the amplification can start from arbitrarily weak seed
fields. This is not the case, however, for possible dynamo mechanisms
where the coupling $B_{{\rm tor}}\to B_{{\rm pol}}$ depends on magnetic
instabilities. Such alternative mechanisms only work if the seed field
is sufficiently strong. There are a variety of magnetic instabilities
on which the dynamo mechanism can be based. \citet{scscfe1996}
started the construction of such models with their on-off dynamo model
where the $\alpha$-effect only exists if the magnetic field exceeds
a certain threshold value determined by the magnetic buoyancy instability.
Other possibilities have been suggested on the basis of both the magnetorotational
instability (MRI) and the Tayler instability \citep{sp2002}.

In this Letter it is demonstrated that in the presence of a shear flow,
including the Hall term in Ohm's law is enough to produce a `shear-Hall
instability' (SHI). The Hall effect produces off-diagonal elements in the
conductivity tensor that are of the first order in the magnetic field. The SHI
is thus a magnetically nonlinear instability, but with the particularity that
there is no threshold value. That is, if the shear is strong enough, the
Hall effect, hence the initial seed field, can be arbitrarily weak.

The natural application of this shear-Hall instability is in newly
born neutron stars. In the first few minutes of its life, the crust
of the star can rotate differentially, as shown in a number of numerical
computations of the supernova core \citep{moemue1989,jamoe1989,difomue2002,kosayasa2004,arbimo2005,budeliot2007}.
Differential rotation in the core
of the collapsing progenitor will induce strong differential rotation
in the newly formed neutron star \citep{otoutobu2005}.

Initially, the star cools down by neutrino emission. As a result of
this cooling, the core becomes distinct from the outer layer, and
the crust condenses at a temperature of a few $10^{9}\,{\rm K}$  
\citep[][and references therein]{dashst2009}. Any differential rotation can
thus only exist in a limited time-interval that depends on the type
of reactions causing the cooling. In the case of the direct URCA process,
the cooling is extremely fast, with estimates giving a characteristic time
of a few minutes \citep{pagewe2006}.

However, the very rapid rotation rates, with periods of order
$10^{-2}\,\rm s$, mean that even just a few minutes is long enough for the star to
undergo several thousand revolutions. Furthermore, we show that the SHI
grows on this very fast rotational timescale, so even though it only
exists for a few minutes, this is sufficient for it to amplify a seed
field by many orders of magnitude. Of course, the time evolution of
the shear is excluded in this simplified model, so the exact amplitude
of the field at the end of the SHI phase must still be deduced from
full nonlinear simulations of the Hall-MHD system.

\section{The shear-Hall instability}

The Hall effect is important whenever a plasma is magnetized enough.  The level
of magnetization is determined by the Hall parameter, $ {\rm R_{B}=\sigma
B_{0}/nec} $, with $ {\rm B_{0}} $ a characteristic magnitude of the field,
${\rm \sigma}$ the electric conductivity, $\rm n$ the electron number density,
${\rm e}$ the electron charge, and ${\rm c}$ the speed of light.  If the Hall
parameter $ {\rm R_{B}} $ is large, the Hall effect is important.

In a neutron star, both the electron number density  and the conductivity vary
with depth. The conductivity also varies with temperature.  To focus on the
basic physics of the instability process, we neglect all such variations and
treat both ${\rm n}$ and ${\rm \sigma}$ as constants. However variations in
${\rm n}$ in particular can interact with the Hall effect in interesting and
unexpected ways \citep{vachol2000,horue2004}.

The magnetic Reynolds number is ${\rm Rm}=R^{2}\,\Omega_{0}/\eta$,
where $R$ is the star's radius, $\Omega_{0}$ the maximal angular
velocity, and $\eta$ its magnetic diffusivity. Scaling length by
$R$, time by $R^{2}/\eta$, $\vec{u}$ by $R\Omega_{0}$, and $\vec{B}$
by $B_{0}$, the induction equation becomes
\begin{eqnarray}
\lefteqn{\frac{\partial\vec{B}}{\partial t}  = }\nonumber\\
&  & -{\rm R_B}\nabla\times((\nabla\times\vec{B})\times\vec{B})
  -\nabla\times(\nabla\times\vec{B})+{\rm Rm}\nabla\times(\vec{u}\times\vec{B}). 
\label{eq:normind} 
\end{eqnarray}
We linearize this equation about the uniform axial field $\vec{B}_{0}=\vec{e}_{z}$,yielding 
\begin{eqnarray}
\lefteqn{\frac{\partial\vec{B}}{\partial t}  =}\nonumber\\
& & -{\rm R_B}\nabla\times((\nabla\times\vec{B})\times\vec{e}_z)
  -\nabla\times(\nabla\times\vec{B})+{\rm Rm}\nabla\times(\vec{u}\times\vec{B}).
\label{eq:linear} 
\end{eqnarray}
The flow $\vec{u}$ is kinematically prescribed as \begin{equation}
\vec{u}=s\Omega\vec{e}_{\phi},\quad{\rm where}\quad\Omega=\frac{1}{\sqrt{1+(s/s_{0})^{2}}},\label{Om}\end{equation}
where $s$ is cylindrical radius, and $s_{0}=0.5$. 
By itself such a flow generates \emph{no} interaction with the imposed
field $\vec{B}_{0}=\vec{e}_{z}$, not even a linearly growing toroidal
field. It is only with the addition of the Hall term that instabilities
may develop in this otherwise stationary basic state.

Investigations of the full field/flow coupling have already been conducted by
\citet{wa1999} and \citet{bate2001} in the local approximation, as well as
globally by \citet{rueki2005}. These works show that the interaction
between the Hall effect and the MRI may lead to both stabilization or
destabilization of the system, depending on the values of the Lundquist number
($ {\rm S= B_{0}R/\sqrt(4\pi\rho)\eta} $, where $\rho$ is the mass density) and
$ {\rm R_{B}} $. In the context of \citet{rueki2005}, our case corresponds to the
limit of small S. For example, Fig. \ref{mm} illustrates how the 
increase in S (the influence of the Lorentz force) has a stabilizing effect.

It is also useful to derive the energy equation associated with either
Eq. (\ref{eq:normind}) or (\ref{eq:linear}). Taking the dot product
of either equation with $\vec{B}$ and integrating over the volume,
one obtains

\begin{equation}
{\frac{{\rm d}}{{\rm d}t}}\int\frac{1}{2}\vec{B}^{2}\, dV=-\int(\nabla\times\vec{B})^{2}\, dV+{\rm Rm}\int\vec{B}\cdot\nabla\times(\vec{u}\times\vec{B})\, dV.\label{energy}
\end{equation}

The Hall term exactly conserves magnetic energy, in both the fully nonlinear
and the linearized versions of the induction equation. The only source of
energy is from the differential rotation. The Hall term is therefore
essentially a catalyst, necessary for the instability to proceed, but
energetically incapable of driving instabilities itself \citep{waho2009}.  This
situation is exactly analogous to the MRI, where the energy source is also
purely from the differential rotation, but a magnetic field is nevertheless
necessary. It is, perhaps, not surprising then that the shear-Hall instability,
like the MRI, grows on the fast rotational timescale, as we demonstrate below.

According to the local analysis of \citet{urrue2005}, for a
general differential rotation depending on both $s$ and $z$, a necessary
condition for the existence of unstable modes is 
\begin{equation}
	(\vec{k}\cdot\vec{B})\left(k_{z}\frac{\partial\Omega}{\partial s}-k_{s}\frac{\partial\Omega}{\partial z}\right)<0,\label{veck}
\end{equation} where $\vec{k}$ is the wavenumber. For $\Omega=\Omega(s)$, this
reduces to 
\begin{equation}
	B_{z}\frac{{\rm d}\Omega}{{\rm d}s}<0,\label{Bzet}
\end{equation}
 so that, for a given gradient of the angular velocity, instability
only occurs for one sign of $B_{z}$ \citep{wa1999,bate2001}. 
For our particular profile (\ref{Om}) this condition is only
fulfilled for positive $B_{z}$, hence ${\rm R_{B}}$. For negative
${\rm R_{B}}$ there is no SHI instability.

Because Eq.~(\ref{eq:linear}) is linear with an axisymmetric
basic state, the different azimuthal modes $\exp(im\phi)$ decouple
and can be studied separately. Each azimuthal mode further decouples
into equatorially symmetric and anti-symmetric modes. We will present
results for S0, A0, S1, and A1, where `S' or `A' refers to the radial
component of the field being equatorially symmetric or anti-symmetric,
and the `0' or `1' refers to the wavenumber $m$.

Equation (\ref{eq:linear}) is solved in a spherical shell using the spherical
harmonics code described by \citet{ho2000}. At both $R_{{\rm in}}=0.7$ and
$R_{{\rm out}}=1$ boundaries, the solution smoothly matches a potential
external field.  At $R_{{\rm out}}$ this is indeed the correct condition, but
at $R_{{\rm in}}$ it is clearly a considerable simplification of the true
physics governing the coupling to an internal core field. Resolutions up to 30
Chebyshev polynomials in $r$ and 70 Legendre functions in $\theta$ were used,
which was sufficient to resolve the eigenmodes up to ${\rm R_{B}}=8$. The
Hall parameter close to unity is appropriate during the initial stages of a
neutron star, and also turns out to be where the shear-Hall instability
operates most efficiently.

\section{Results}

\subsection{Stability maps}

\begin{figure}[htb]
\includegraphics[clip,width=0.5\textwidth]{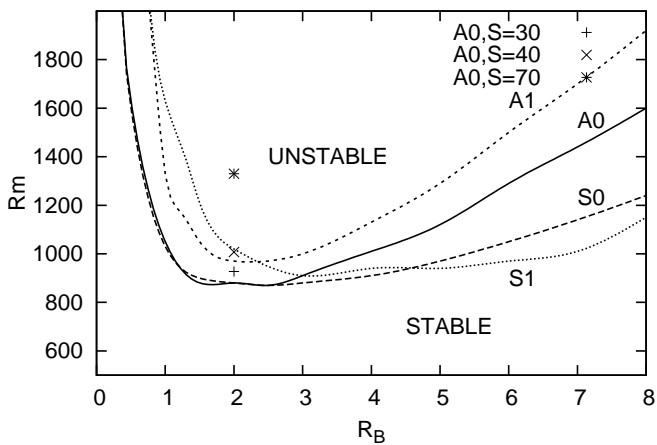} 

\caption{Stability curves for the axisymmetric (S0 and A0) and non-axisymmetric
(S1 and A1) modes. The single points belong to the system where the flow
equation was treated in addition to the induction equation.}\label{mm} \par\end{figure}

Figure \ref{mm} shows the instability curves for the four modes S0,
A0, S1, and A1. Over most of the ${\rm R_{B}}$ range the axisymmetric
modes are slightly preferred over the non-axisymmetric ones, but the
curves are otherwise all qualitatively similar. The minimum value
of ${\rm Rm}$ occurs when ${\rm R_{B}\simeq2-4}$. At this minimum
the shear frequency $\omega_{{\rm sh}}$ is around 400 times the Hall
frequency $\omega_{{\rm H}}$, consistent with the result of \citet{urrue2005}
that the instability only exists for $|\omega_{{\rm sh}}|>|\omega_{{\rm H}}|$.

If ${\rm R_{B}}$ is smaller than $\simeq2-4$ the system  tends
toward the (stable) pure differential rotation case, thus explaining
the asymptote ${\rm Rm}\to\infty$ as ${\rm R_{B}}\to0$. At the other
end, as ${\rm R_{B}}$ becomes large the Hall effect increasingly
dominates, but as noted above, it is energetically incapable of driving
any instability itself, so ${\rm Rm}$ must again increase to compensate.
For both ${\rm R_{B}}\ll1$ and ${\rm R_{B}}\gg1$, therefore, the differential
rotation must become increasingly strong for the instability to operate.

An intriguing feature of the small ${\rm R_{B}}$ ends of all four
instability curves is that none of them shows any evidence of tilting
back over to the right (down to our lowest investigated $\rm R_{B}$ value
of 0.4), that is, there is no indication of the instability
switching off again for sufficiently large ${\rm Rm}$. This is perhaps not so surprising for the axisymmetric
modes; For example, the classical, axisymmetric
MRI has the same feature. For
nonaxisymmetric modes, though, it is often the case that a
strong enough differential rotation suppresses them again 
\citep[e.g.][]{ruehogesc2007,hoterue2010}. It is not entirely clear why
this does not occur here -- or perhaps it does, but only at much higher values
of ${\rm Rm}$  than what we were able to compute here.

It is also interesting that there is no region where a single mode
dominates: both A and S modes, and both axisymmetric and non-axisymmetric ones,
are excited at comparable parameter values. This suggests that the
fully nonlinear regime  might consist of a mixture of several
modes, which can also be considered as an `oblique rotator'. If different
modes are indeed present simultaneously, the field configuration and
amplitude in the two hemispheres would necessarily be different. If
the heat transport is furthermore dominated by the magnetic suppression
of the thermal conductivity tensor \citep{sc1990,hehe2001}, then 
the Hall effect would thereby lead to different temperatures,
and hence X-ray emission \citep{bepa2002}, in the two
hemispheres.

\subsection{Growth rates}

\begin{figure}
\noindent \begin{centering}
\includegraphics[clip,width=0.5\textwidth]{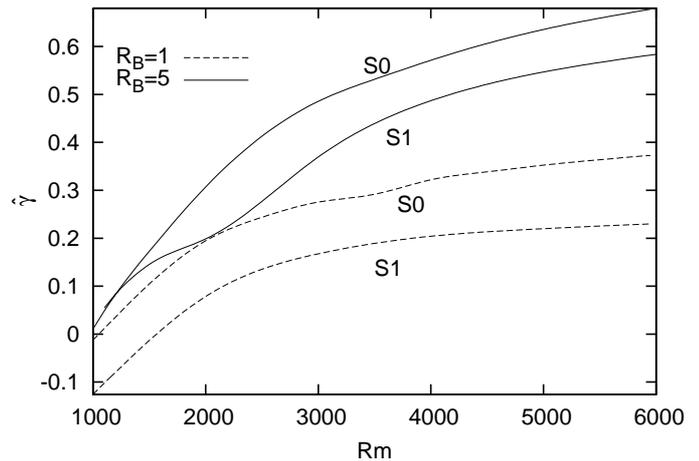}
\end{centering}

\caption{$\hat{\gamma}$ is a nondimensionalized growth rate ($\hat{\gamma}=\tau_{\rm rot}\gamma$, where $\gamma$ is in physical units), 
of the S0 and S1 modes, for ${\rm R_{B}=1}$ and $5$. The profiles for A0 and A1 are very similar, and are not
shown here.}

\label{gr} 
\par\end{figure}

To assess the possible astrophysical significance of the
shear-Hall instability, we need to know not only for what parameter
values it exists, but also what its associated growth rates are. Figure
\ref{gr} shows the growth rate scaled on the rotational timescale.
(That is, if $\gamma_{\rm diff}$ is the growth rate on the chosen diffusive
timescale $R^{2}/\eta$, then $\hat{\gamma}=(2\pi/\rm Rm)\gamma_{\rm diff}$ is the
growth rate on the rotational timescale $\tau_{\rm rot}=2\pi/\Omega_{0}$.) For both
${\rm R_{B}=1}$ and ${\rm R_{B}=5}$, for large enough ${\rm Rm}$
these growth rates $\hat{\gamma}$ eventually saturate at $\rm O(1)$
values. As before, for the initial onset, all four modes behave
comparably, although the axisymmetric modes do have higher
growth rates. There is no preferred equatorial symmetry, because the S and
A modes are very similar. Surprisingly, even though the stability
map shows that a nonaxisymmetric mode S1 is less stable in the $\rm R_{B}$ 
region from 4 onward, its growth is only stronger near the stability line.
For higher Rm, the axisymmetric modes are more dominant.

Of the three timescales that are `obviously' present in the problem,
namely the fast rotational timescale $\tau_{{\rm rot}}$,
the much slower diffusive $\tau_{{\rm diff}}=R^{2}/\eta=(\rm Rm/2\pi)\tau_{{\rm rot}}$,
and Hall $\tau_{{\rm Hall}}=\tau_{{\rm diff}}/{\rm R_{B}}$ timescales,
the shear-Hall instability thus grows the fastest, and is independent
of both $\tau_{{\rm diff}}$ and $\tau_{{\rm Hall}}$. This is surprising, 
given that it does clearly depend on both the shear-effect and the Hall effect. 
On the other hand, as noted above, its energy source is entirely drawn 
from the differential rotation. In this regard it
is quite different from a classical $\alpha\omega$ dynamo, for example,
where the growth rate is the geometric mean of $\tau_{{\rm rot}}$
and $\tau_{\alpha}$, but both the differential rotation and $\alpha$
also contribute to the energy balance.

\subsection{Wave numbers}

\begin{figure}[htb]
\noindent \begin{centering}
\includegraphics[clip,width=0.5\textwidth]{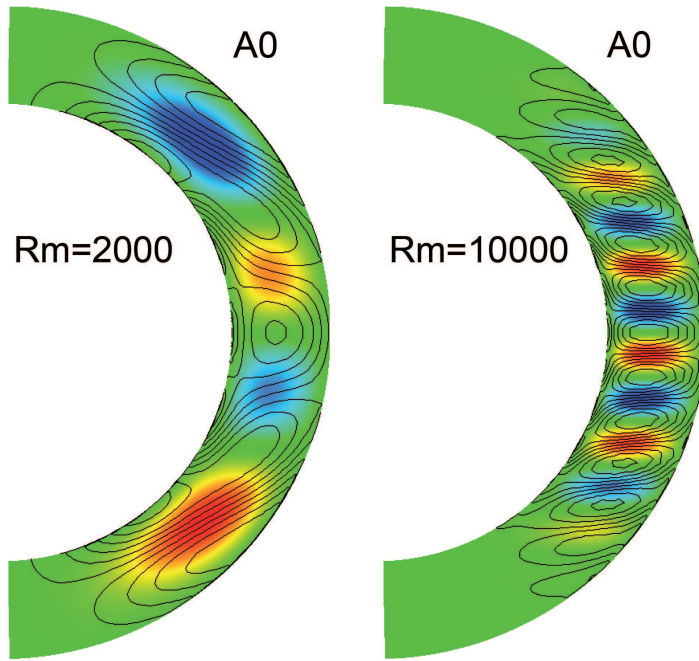}
\par\end{centering}

\caption{The axisymmetric solution A0 for ${\rm R_{B}=1}$, and ${\rm Rm=2000}$
on the left and ${\rm Rm=10000}$ on the right. The other modes have
the same pattern of exhibiting finer structure in $z$ for larger
${\rm Rm}$.}

\label{ff} 
\end{figure}

The spatial properties of the eigensolutions are given in Fig. \ref{ff}.  The
color map represents the toroidal field structure, while the contours depict the
field lines of the poloidal component. Comparing the two plots, at ${\rm
Rm=2000}$ and $10000$, suggests a transition to smaller scales as ${\rm Rm}$
increases. The wavenumber also increases for (turbulent) mean-field dynamos
at increasingly supercritical dynamo numbers.  However, thanks to the high values
of their eddy diffusivity, turbulent dynamos operate with moderate values of
${\rm Rm}$. The SHI though may easily operate with very high ${\rm Rm}$, and
may play a dominant role in driving MHD turbulence if ${\rm Rm\gg1}$.

The relative amplitudes of the toroidal field never exceed that of
the poloidal field by more than one order of magnitude. Independent
of the ratio of the Hall time and the rotation time, the coupling
between the components via the Hall term enables their simultaneous,
and comparable, amplification.

\section{Discussion}

It remains to be seen to what extent the SHI is important in
the regime of newborn neutron stars. We first need to establish
the reasonable values for the Hall parameter and the
magnetic Reynolds number.  It turns out that owing to a low diffusivity
($\eta=10^{-(2..3)}\, \mbox{\rm{cm}}^{2}\,{\rm{s}}^{-1}$) \citep[as treated in the code of][]{po2008}, the
microphysical ${\rm Rm}$ is orders of magnitude above the stability line 
in Fig. \ref{mm}.  If the effects of turbulence
are important, according to \citet{narebopa2008} diffusivity is much
larger and the corresponding ${\rm Rm}$ is lower ($\sim10^{3}$), but
still supercritical.

The onset of the shear-Hall instability requires a seed field of the
magnitude determined by the actual ${\rm Rm}$. This field can be provided
by various other proposed mechanisms, from flux-conservation to dynamo
processes \citep[for an overview, see][]{sp2009}.  The turbulent ${\rm Rm}$
would require ${\rm R_{B}} \gtrsim 1$, which, for the temperatures of
$10^{9\dots10}\, \rm K$ and densities of $10^{10\dots13}\, \rm{g}\, \mbox{\rm cm}^{-3}$,
corresponds to $B\gtrsim 10^{12}\, \rm G$.  If no turbulence is present, the
seed field can be much smaller provided that the stability line remains
univalued when $\rm {R_{B}}$ tends to zero.

In the situation where the magnetic fields are strong, there is a coupling
between the fields and the differentially rotating flow, which leads to a
stable state of uniform rotation. The timescale for this process is 
$\tau_{\rm A}=R\sqrt{4\pi\rho}/B$, the Alfv\'{e}n time \citep{sh2000}, which is only
$\sim10\,\rm s$ for the neutron star conditions. Is the SHI quick enough to be
of any significance?  Taking the fast rotation of neutron stars into account ,
our results (Fig. \ref{gr}) show that the field can easily amplify several
orders of magnitude within the constraining time interval.

Because we are convinced that the instability conditions are fulfilled and the
growth is quick enough, there is only the field geometry left to analyze
in the context of the standard, inclined magnetic dipole model, which is used to
explain the spin-down of pulsars. The SHI generates, as shown, small-scale 
fields for large $\rm Rm$. The presence of turbulence is,
it seems,  a necessary requirement for producing a global field via
this instability. On the other hand, X-ray spectra of pulsars sometimes
indicate an additional, strong small-scale surface component \citep{pazasatr2002,sapazate2002}.
It is possible that the SHI is involved in generating the small-scale 
structure, while the large-scale component should be attributed to another process.

Rapid growth, as well as the easy excitation, indicates that the shear-Hall
instability may play an important role in the evolution of the field of newborn
neutron stars, prompting further, nonlinear analysis of the field/flow
coupling and raising questions about its interaction with other
field-generating mechanisms.

\bibliographystyle{aa} 
\bibliography{krh_SHI_ph} 
\end{document}